\documentclass[aps,preprint,showpacs, showkeys]{revtex4}%
\usepackage{amsfonts}
\usepackage{amsmath}
\usepackage{amssymb}
\usepackage{bm}
\usepackage{graphicx}%
\providecommand{\U}[1]{\protect \rule{.1in}{.1in}}

\makeatletter

\newcommand{\Rmnum}[1]{\expandafter \@slowromancap \romannumeral #1@}
\makeatother
\begin{document}
\title{Non-perturbative tests for the Asymptotic Freedom in the $\mathcal{PT}%
$-symmetric $(-\phi^{4})_{3+1}$ theory}
\author{${}^{1}$ Abouzeid Shalaby\footnote{email:~amshalab@mans.edu.eg}}
\author{${}^{2}$ Suleiman S. Al-Thoyaib}
\affiliation{${}^{1}$Physics Department, Faculty of Science, Mansoura University, Egypt\\ ${}^{2}$ Physics Department, Faculty of Science, Qassim University, Saudi Arabia}

\keywords{Effective potential, Hierarchy problem, Cosmological constant, non-Hermitian
models, $\mathcal{PT}$-symmetric theories, Quasi-Hermitian field theory.}
\pacs{12.60.Cn, 98.80.Cq, 12.90.+b}

\begin{abstract}
In the literature, the asymptotic freedom property of the $(-\phi^{4})$ theory
is always concluded from real-line calculations while the theory is known to
be a non-real-line one. In this article, we test the existence of the
asymptotic freedom in the $(-\phi^{4})_{3+1}$ theory using mean field
approach. In this approach and contrary to the original Hamiltonian, the
obtained effective Hamiltonian is rather a real-line one. Accordingly, this
work resembles the first reasonable analysis for the existence of the
asymptotic freedom property in the $\mathcal{PT}$-symmetric $(-\phi^{4})$
theory. In this respect, we calculated three different amplitudes of different
positive dimensions (in mass units) and find that all of them goes to very
small values at high energy scales (small coupling) in agreement with the
spirit of the asymptotic freedom property of the theory. To test the validity
of our calculations, we obtained the asymptotic behavior of the vacuum
condensate in terms of the coupling, analytically, and found that the
controlling factor $\Lambda$ has the value $\frac{(4 \pi)^{2}}{6}= 26. 319$
compared to the result $\Lambda=26.3209$ from the literature which was
obtained via numerical predictions. We assert that the non-blow up of the
massive quantities at high energy scales predicted in this work strongly
suggests the possibility of the solution of the famous hierarchy puzzle in a
standard model with $\mathcal{PT}$-symmetric Higgs mechanism.

\end{abstract}
\maketitle

One of the biggest puzzles in the theory of particle interactions is the
Hierarchy problem \cite{DJOUADI}. Due to this conceptual problem, all
quantities of positive mass dimensions in the scalar sector of the Standard
model (e.g. the Higgs mass) blow up to unacceptable values at high energy
scales. The worst manifestation of the Hierarchy problem is in the vacuum
energy as it behaves like $\mu^{4}$ with $\mu$ is a unit mass. This leads to
the most unacceptable discrepancy between theory and experiment in the
prediction of the cosmological constant (vacuum energy). Indeed, the root of
the Hierarchy problem stems from the fact that the scalar Higgs mechanism
played by the Hermitian $\phi^{4}$ theory has a positive Beta function which
up to second order in the $\phi^{4}$ coupling $g$ is given by;
\[
\beta \left(  g\right)  =\frac{3g^{2}}{\left(  4\pi \right)  ^{2}} \mathtt{.}
\]
One can easily show that the positiveness of the Beta function of the
$\phi^{4}$ theory \ will lead to a huge Higgs mass at high energy scales
\cite{abohir}. By catching the main reason for the Hierarchy problem, one may
wonder if the existence of a scalar theory with rather negative Beta function
(i.e. asymptotically free) will help in solving the Hierarchy problem in the
standard model of particle interactions?. In Ref. \cite{abohir}, we argued
that such reasoning becomes legitimate since the discovery of the physical
acceptability of non-Hermitian and $\mathcal{PT}$-symmetric theories
\cite{bendh1,abohir,abophi41p1,abophi61p1,zadah,bendr,bend2005,bendg1}.

In spite of the beauty of the idea of employing a now physically acceptable
theory with bounded from above potential to play the role of Higgs mechanism,
in the literature, all the claims about the asymptotic freedom property of the
$\mathcal{PT}$-symmetric ($-\phi^{4}$) scalar field theory were built on a
real-line calculation (inaccurate) while the theory is well known to be a
non-real line one \cite{Symanzik,bendf,Frieder}. In view of this realization,
our aim in this work is to test the existence of the more than important
asymptotic freedom property for $\mathcal{PT}$-symmetric ($-\phi^{4}$) scalar
field theory but this time using algorithms which proved to be reliable for
the study of non-real line problems.

To shed light on how important to employ an asymptotically free $\mathcal{PT}%
$-symmetric ($-\phi^{4}$) scalar field theory to play the role of the Higgs
mechanism in the standard model, we mention some possible problems existing in
the standard model and its extensions. In the standard model, the spontaneous
symmetry breaking adds a large shift to the vacuum energy of the form
$\Delta \langle0|H|0\rangle \sim-CB^{4}$, where $H$ \ is the Hamiltonian
operator, $C$ is dimensionless and $B$ is the vacuum condensate \cite{Peskin}.
This shift is finite but still large. As we will see in this work, contrary to
the corresponding Hermitian theory, the vacuum energy of the $\mathcal{PT}%
$-symmetric ($-\phi^{4}$) scalar field theory is tiny for all energy scales
and thus gives a clue to benefits that may be drawn by its employment in the
standard model. This discrepancy between the features of the two theories
(Hermitian and the non-Hermitian theories) can be understood from the fact
that for a theory with negative Beta function, the coupling will be dragged to
the origin at high energy scales \cite{kaku} which in turn drag all the
dimnesionfull quantities to small values while the reverse is correct for a
theory with positive Beta function. This is the origin of the gauge Hierarchy
problem. For the solution of this problem, different algorithms have been
introduced. For instance, in the SUSY regime there exists natural cancellation
in the dimensionful parameters that turned those parameters protected against
perturbations even for very high energy scales \cite{susy}. However, SUSY
introduces an upper limit to the Higgs mass by 130 $GeV$ and some of its mass
spectra are of one $TeV$ which expose this theory direct to the fire of the
LHC experiments. Another algorithm for the solution of the Hierarchy problem
is to consider the Higgs particle as a composite state bounded by a new set of
interactions (Technicolors) \cite{tecni}. However, the technicolor model is
strongly constrained from precision tests of electroweak theory at LEP and SLC
experiments \cite{tecni1} . Also, this algorithm has mass spectra of about one
$TeV$ and it is under the direct test of the LHC experiments. Once more, a
recent algorithm is suggested for which the $SU(2)\times U(1)$ symmetry is
broken via the compactification of an extra dimension \cite{extra}. In fact,
particles in this model attain masses through the expectation value of the
fifth ( for instance) component of the gauge field. To some physicists, the
digestion of extra dimension is not that easy and can be accepted to them at
most as a mathematical modeling to the problem.

The introduction of a new scenario that may overcome the Hierarchy problem as
well as does not introduce extra problems may be possible. Indeed, the scalar
field in the standard model is the source of the hierarchy problem and thus it
would be very important to have a scalar field with unproblematic features. In
this work we, study the flow of different quantities of positive mass
dimension in the $\mathcal{PT}$-symmetric $(-\phi^{4})$ theory and show that
they do not blow up at high energy scales which is the reverse of the behavior
of the Hermitian $\phi^{4}$ at those scales. Although our calculations are
carried out for a one component field while the one used in the standard model
has a higher group structure, it is not expected that the group structure will
change the amazing asymptotic freedom property of the theory which means that
the results in our work strongly recommends the replacement of the
conventional Higgs mechanism by a $\mathcal{PT}$-symmetric one which then is
expected to overcome the Hierarchy problem.

The attractive idea of a possible safe employment of the $\mathcal{PT}%
$-symmetric Higgs mechanism is in fact confronted by two main technical
problems. The first problem is the believe of the lack of a reliable
calculational algorithm to follow for $\mathcal{PT}$-symmetric ($-\phi^{4}$)
field theory. Among the well known non-perturbative methods that works for
$\mathcal{PT}$-symmetric theories is the complex contour treatment applied
successfully for the $0+1$ space-time dimensions (quantum mechanics)
\cite{bendr}. However, this method is not willing to be applicable for higher
space-time dimensions. Regarding this problem, in a previous work
\  \cite{aboeff}, we discovered (for the first time) that the mean field
approach which is famous in field theory calculations works well for
$\mathcal{PT}$-symmetric ($-\phi^{4}$) theory. While its applicability in
higher space-time dimensions is not questionable, in Ref. \cite{aboeff}, we
exposed the effective field approach to a quantitative test and showed that it
is accurate even at the level of first order in the coupling which means that
the first problem has been solved. The second problem that is confronting the
progress in the study of $\mathcal{PT}$-symmetric ($-\phi^{4}$) theory in the
real world of $3+1$ space time dimensions is that the metric operator for this
theory is very hard to be obtained. However, in the mean field regime in
Ref.\cite{aboeff}, we realized that the propagator of the $\mathcal{PT}%
$-symmetric ($-\phi^{4}$) theory has the correct sign. Recently, Jones
\textit{et.al} conjectured that mean field theory may know about the metric.
In fact, Jones et.al. showed that the Greens functions in the mean field
approach are taking into account the employment of the metric operator
\cite{jonesgr,jonesgr1}. In view of these explanations, we think that a
non-problematic $\mathcal{PT}$-symmetric Higgs mechanism is now possible and
it is just a matter of known calculations. In fact, the version of mean field
approach used by one of us in Ref. \cite{aboeff} mimics the way of breaking
the symmetry in the standard model and thus we assert that it is the most
known plausible method to use for the study of the $\mathcal{PT}$-symmetric
Higgs mechanism. However, as a first step toward the employment of the
$\mathcal{PT}$-symmetric Higgs mechanism we need to check the existence of the
asymptotic freedom property in the prototype example of the one component
$\mathcal{PT}$-symmetric $(-\phi^{4})$ theory using the mean field approach.

The motivation behind the application of the mean field approach for the
$\mathcal{PT}$-symmetric $(-\phi^{4})$ theory is that, in the literature, all
the claims about the asymptotic freedom property of the $\mathcal{PT}%
$-symmetric ($-\phi^{4}$) scalar field theory were built on a real-line
calculation while the theory is well known to be a non-real line one. In view
of this realization, our main target in this work is to assure the existence
of the asymptotic freedom property for $\mathcal{PT}$-symmetric ($-\phi^{4}$)
scalar field theory but this time using the reliable mean field approach (as
used by us in Ref. \cite{aboeff}) which implements the use of the metric as
well. Although our work stresses the one component $\mathcal{PT}$-symmetric
($-\phi^{4}$) scalar field theory, its extension to charged $\mathcal{PT}%
$-symmetric ($-\phi^{4}$) scalar field theory is direct.

Since the calculational algorithm we will follow in this work is the mean
field approach as presented in our work in Ref.\cite{aboeff}, as a reminder,
we summarize its results. To do that, we consider the Lagrangian density of
the massless $\mathcal{PT}$-symmetric ($-\phi^{4}$) scalar field theory;
\[
H(x)=\frac{1}{2}\left(  \left(  \nabla \phi(x)\right)  ^{2}+\pi^{2}(x)\right)
-\frac{g}{2}\phi^{4}(x),
\]
where $\phi(x)$ is the field variable, $\pi(x)$ is the canonical conjugate
momentum field and $g$ is the coupling constant. The effective field approach
uses the application of a canonical transformations of the form;
\[
\phi=\psi+B,\text{ \ }\pi=\Pi=\dot{\psi},\text{\  \  \ }%
\]
with $B$ is a vacuum condensate and $\psi$ is a fluctuating field. Thus one
obtains the form
\[
H=H_{0}+H_{I},
\]
where
\begin{align}
H_{0}  &  =\frac{1}{2}\left(  \left(  \nabla \psi \right)  ^{2}+\Pi^{2}%
+M^{2}\psi^{2}\right)  ,\nonumber \\
\ H_{I}  &  =-\frac{g}{2}\left(  \psi^{4}+4B\psi^{3}\right)  +\left(
-\frac{1}{2}M^{2}-3gB^{2}\right)  \psi^{2}-2gB^{3}\psi. \label{effd1}%
\end{align}
and $M$ is the mass of the field $\psi$.

We used the known relations of the effective potential of the form;%
\begin{equation}
\frac{\partial V_{eff}}{\partial B}=0\text{, }\frac{\partial^{2}V_{eff}%
}{\partial B^{2}}=M^{2}, \label{paramf}%
\end{equation}
where $V_{eff}=\langle0|H|0\rangle$ and $M$ is the mass of the $\psi$ field.
Up to first order in the coupling and in $0+1$ dimensions, we were able to
obtain the equations;
\begin{align}
\left(  -2g\right)  B^{3}+\left(  -\frac{3}{M}g\right)  B  &  =0,\nonumber \\
\allowbreak \left(  -6g\right)  B^{2}-\frac{3}{M}g\allowbreak &  =M^{2}.
\end{align}
For $B\neq0$, one can get the parametrization
\begin{align}
B  &  =-\sqrt{\frac{M^{2}}{-4g}},\nonumber \\
M  &  =\sqrt[3]{6g}. \label{param}%
\end{align}
Since $B$ is imaginary, the effective Hamiltonian in Eq.(\ref{effd1}) is
non-Hermitian and $\mathcal{PT}$-symmetric. Moreover, while the original
theory is a non-real line one, the effective form is a real-line theory
\cite{quart,realline}. In fact, this is a very important realization since the
real-line effective field calculations can be extended to higher dimensional
cases (field theory) for which non-real line problems can not be treated using
the complex contour method.

Note that the results presented in Ref.\cite{jonesgr} coincides with our
(older) above results. Besides, the conditions $\frac{\partial V_{eff}%
}{\partial B}=0$, $\frac{\partial^{2}V_{eff}}{\partial B^{2}}=M^{2},$ we used
in Ref.\cite{aboeff} are coincident with the variational conditions
$\frac{\partial V_{eff}}{\partial B}=0$, $\frac{\partial^{2}V_{eff}}{\partial
M}=0$. In fact, this method not only gives accurate results for the energy
spectra and the condensate but also results in the correct sign of the
propagator (no ghosts) which is assured by the positiveness of the $M^{2}$
parameter. This result is a clue for the possibility of the disappearance of
the metric operator from the calculation of the Greens functions in the
effective field approach. In fact, the assertion that mean field approach
knows about the metric has been clearly proved by Jones \textit{et.al }$\ $in
Refs \cite{jonesgr,jonesgr1}.

According to the above results, the effective field approach applied by one of
us for the first time in Ref.\cite{aboeff} is a satisfactory algorithm for the
calculations in the $\mathcal{PT}$-symmetric ($-\phi^{4}$) scalar field theory
which does not need the difficult calculation of the metric operator or the
may be impossible complex contour integrations followed in the quantum
mechanical case.

Unlike the quantum mechanical case, in quantum field theory one always
confronted by infinities in the amplitude calculations. Since we restrict
ourself to first order calculations, normal ordering is a valuable tool to
exclude infinities at this order of calculations. In this tool, the first
order vacuum energy can be obtained by normal ordering the operators in the
effective Hamiltonian with respect to two different mass scales
\cite{Coleman,efbook}. The benefit of this method is that it accounts not only
for first order diagrams but also for all the higher order cactus diagrams
\cite{wen-Fa, changcac}. Moreover, in Ref.\cite{mag}, it has been shown that
the method is equivalent to the first order calculations with the
regularization carried out via the addition of a counter term.

To start the algorithm, consider a normal-ordered Hamiltonian density with
respect to a mass parameter $m$ of the form;%
\begin{equation}
H=N_{m}\left(  \frac{1}{2}\left(  \left(  \nabla \phi \right)  ^{2}+\pi
^{2}+m^{2}\phi^{2}\right)  -\frac{g}{4!}\phi^{4}\right)  . \label{hamilt2}%
\end{equation}
We can use the relation \cite{Coleman,normal2}
\begin{equation}
N_{m}\exp \left(  i\beta \phi \right)  =\exp \left(  -\frac{1}{2}\beta^{2}%
\Delta \right)  N_{M=t\cdot m}\exp \left(  i\beta \phi \right)  \text{,}
\label{normal2}%
\end{equation}
with
\begin{equation}
\Delta=-\langle0\left \vert \phi(x)\phi(x)\right \vert 0\rangle,
\end{equation}
to rewrite the Hamiltonian normal ordered\ with respect to a new mass
parameter $M=\sqrt{t}\cdot m$. In eq.(\ref{normal2}), expanding both sides and
equating the coefficients of the same power in $\beta$ yields the result;%

\begin{align}
N_{m}\phi &  =N_{M}\phi,\nonumber \\
N_{m}\phi^{2}  &  =N_{M}^{2}\phi^{2}+\Delta,\nonumber \\
N_{m}\phi^{3}  &  =N_{M}\phi^{3}+3\Delta N_{M}\phi,\label{normall}\\
N_{m}\phi^{4}  &  =N_{M}\phi^{4}+6\Delta N_{M}\phi^{2}+3\Delta^{2},\nonumber
\end{align}
Also, it is easy to obtain the result \cite{Coleman}%
\begin{equation}
N_{m}\left(  \frac{1}{2}\left(  \nabla \phi \right)  ^{2}+\frac{1}{2}\pi
^{2}\right)  =N_{M}\left(  \frac{1}{2}\left(  \nabla \phi \right)  ^{2}+\frac
{1}{2}\pi^{2}\right)  +E_{0}(M)-E_{0}(m), \label{norkin1}%
\end{equation}
where
\[
E_{0}(\Omega)=\frac{1}{4}\int \frac{d^{3}k}{\left(  2\pi \right)  ^{3}}\left(
\frac{2k^{2}+\Omega^{2}}{\sqrt{k^{2}+\Omega^{2}}}\right)  =I_{1}+I_{2},
\]
with
\begin{equation}
I_{1}=\frac{1}{2}\int \frac{d^{3}p}{\left(  2\pi \right)  ^{3}}\frac{p^{2}%
}{\sqrt{p^{2}+m^{2}}}=\frac{1}{2}\frac{1}{\left(  4\pi \right)  ^{\frac{3}{2}}%
}\frac{d}{2}\left(  \frac{\Gamma \left(  \frac{1}{2}-\frac{d}{2}-1\right)
}{\Gamma \left(  \frac{1}{2}\right)  }\left(  \frac{1}{m^{2}}\right)
^{\frac{1}{2}-\frac{d}{2}-1}\right)  ,
\end{equation}%
\begin{equation}
I_{2}=\frac{1}{4}\int \frac{d^{3}p}{\left(  2\pi \right)  ^{3}}\frac{m^{2}%
}{\sqrt{p^{2}+m^{2}}}=\frac{m^{2}}{4}\frac{1}{\left(  4\pi \right)  ^{\frac
{3}{2}}}\left(  \frac{\Gamma \left(  \frac{1}{2}-\frac{d}{2}\right)  }%
{\Gamma \left(  \frac{1}{2}\right)  }\left(  \frac{1}{m^{2}}\right)  ^{\frac
{1}{2}-\frac{d}{2}}\right)  .
\end{equation}
Here, $\Gamma$ is the gamma function and$\ d$ is the spatial dimension. Accordingly,%

\begin{equation}
\Delta E_{0}=\left(
\begin{array}
[c]{c}%
\frac{3}{8\sqrt{\pi}}\frac{M^{4}}{\epsilon \left(  4\pi \right)  ^{\frac{3}{2}}%
}-\frac{1}{2\sqrt{\pi}}M^{4}\left(  \left(  \frac{\frac{3}{4}\gamma-\frac
{5}{8}}{\left(  4\pi \right)  ^{\frac{3}{2}}}-\frac{3}{4}\frac{\ln4\pi}{\left(
4\pi \right)  ^{\frac{3}{2}}}\right)  -\frac{3}{4}m^{4}\frac{\ln t}{\left(
4\pi \right)  ^{\frac{3}{2}}}\right)  +\allowbreak O\left(  \epsilon \right)  \\
+\frac{3}{8\sqrt{\pi}}\frac{m^{4}}{\epsilon \left(  4\pi \right)  ^{\frac{3}{2}%
}}-\frac{1}{2\sqrt{\pi}}\left(  m^{4}\left(  \frac{\frac{3}{4}\gamma-\frac
{5}{8}}{\left(  4\pi \right)  ^{\frac{3}{2}}}-\frac{3}{4}\frac{\ln4\pi}{\left(
4\pi \right)  ^{\frac{3}{2}}}\right)  -\frac{3}{4}m^{4}\frac{\ln t}{\left(
4\pi \right)  ^{\frac{3}{2}}}\right)  +\allowbreak O\left(  \epsilon \right)  \\
-\frac{1}{4\sqrt{\pi}}\frac{M^{4}}{\epsilon \left(  4\pi \right)  ^{\frac{3}{2}%
}}-\frac{1}{4\sqrt{\pi}}M^{4}\left(  \left(  \frac{\ln4\pi}{\left(
4\pi \right)  ^{\frac{3}{2}}}-\frac{\gamma-1}{\left(  4\pi \right)  ^{\frac
{3}{2}}}\right)  +\frac{\ln t}{\left(  4\pi \right)  ^{\frac{3}{2}}}\right)
+\allowbreak O\left(  \epsilon \right)  \\
-\left(  -\frac{1}{4\sqrt{\pi}}\frac{m^{4}}{\epsilon \left(  4\pi \right)
^{\frac{3}{2}}}-\frac{1}{4\sqrt{\pi}}m^{4}\left(  \left(  \frac{\ln4\pi
}{\left(  4\pi \right)  ^{\frac{3}{2}}}-\frac{\gamma-1}{\left(  4\pi \right)
^{\frac{3}{2}}}\right)  +\frac{\ln t}{\left(  4\pi \right)  ^{\frac{3}{2}}%
}\right)  \right)  +\allowbreak O\left(  \epsilon \right)
\end{array}
\right)  ,
\end{equation}
where $\epsilon=\frac{3-d}{2}$, $\Delta E_{0}=E_{0}(M)-E_{0}(m)$, $\gamma$ is
the Euler number given by
\[
\gamma=\lim \limits_{n\rightarrow \infty}\left(  \sum \limits_{m=1}^{n}\frac
{1}{m}-\ln n\right)  ,
\]
and
\[
t=\frac{\mu^{2}}{m^{2}}=\frac{\nu^{2}}{m^{2}}.
\]
Here $\mu$ and $\nu$ are unit masses chosen to make the argument of the
logarithm dimensionless. Also, the relation $\frac{\mu^{2}}{m^{2}}=\frac
{\nu^{2}}{m^{2}}$ has been employed to fix the renormalization scheme
\cite{collinbook}. As $\epsilon \rightarrow0$, $\Delta E_{0}$ can be simplified
as;
\[
\Delta E_{0}=\frac{1}{64\pi^{2}}\left(  M^{4}-m^{4}\right)  \left(
1-\gamma+\ln4\pi+\ln t\right)  .
\]

The mass shift $m\rightarrow M$ should be accompanied by the canonical
transformation; \cite{efbook}%

\begin{equation}
\left(  \phi,\pi \right)  \rightarrow \left(  \psi+B,\Pi \right)  .
\end{equation}
The field\ $\psi$ has mass $M=\sqrt{t}\cdot m$, $B$ is a constant, the field
condensate, and $\Pi$ is the conjugate momentum ($\overset{\cdot}{\psi}$).
Therefore, the Hamiltonian in Eq.(\ref{hamilt2}) can be written in the form;

\bigskip%
\begin{equation}
H=\bar{H}_{0}+\bar{H}_{I}+\bar{H}_{1}+E, \label{quasi}%
\end{equation}
where
\begin{align*}
\bar{H}_{0}  &  =N_{M}\left(  \frac{1}{2}\left(  \Pi^{2}+\left(
\triangledown \psi \right)  ^{2}\right)  \right)  +\frac{1}{2}N_{M}\left(
m^{2}-\frac{g}{2}\left(  B^{2}+\Delta \right)  \right)  \psi^{2},\\
\bar{H}_{I}  &  =\frac{-g}{4!}N_{M}\left(  \psi^{4}+4B\psi^{3}\right)
\end{align*}
$\bar{H_{1}}$ can be found as
\begin{equation}
\bar{H}_{1}=N_{M}\left(  m^{2}-\frac{g}{4!}\left(  4B^{2}+3\Delta \right)
\right)  B\psi,
\end{equation}
and the field independent terms can be regrouped as;
\begin{equation}
E=\frac{1}{2}\left(  m^{2}-\frac{12g\Delta}{4!}\right)  B^{2}-\frac{g}%
{4!}B^{4}+\Delta E_{0}-\frac{3g\Delta^{2}}{4!}+\frac{1}{2}m^{2}\Delta
.\label{e3p1}%
\end{equation}

Taking $d=3-2\epsilon$, we get%
\begin{align*}
\Delta &  =\frac{1}{\left(  4\pi \right)  ^{\frac{d+1}{2}}}\left(  \frac
{\Gamma \left(  1-\frac{d+1}{2}\right)  }{\left(  M^{2}\right)  ^{1-\frac
{d+1}{2}}}\right)  -\frac{1}{\left(  4\pi \right)  ^{\frac{d+1}{2}}}\left(
\frac{\Gamma \left(  1-\frac{d+1}{2}\right)  }{\left(  m^{2}\right)
^{1-\frac{d+1}{2}}}\right)  ,\\
&  =\frac{1}{16\pi^{2}}\allowbreak \left(  M^{2}-m^{2}\right)  \left(
\gamma-1-\ln4\pi+\ln t\right)  ,
\end{align*}
Substituting for $\Delta E_{0}$ and $\Delta$ in Eq.(\ref{e3p1})we get;%

\begin{align}
E  &  =\frac{1}{2}\left(  m^{2}-\frac{12g}{4!}\frac{1}{16\pi^{2}}%
\allowbreak \left(  M^{2}-m^{2}\right)  \left(  \gamma-1-\ln4\pi+\ln t\right)
\right)  B^{2}\nonumber \\
&  -\frac{g}{4!}B^{4}+\left(  \frac{1}{64\pi^{2}}\left(  M^{4}-m^{4}\right)
\left(  1-\gamma+\ln4\pi+\ln t\right)  \right) \nonumber \\
&  -\frac{3g\left(  \frac{1}{16\pi^{2}}\allowbreak \left(  M^{2}-m^{2}\right)
\left(  \gamma-1-\ln4\pi+\ln t\right)  \right)  ^{2}}{4!}\\
&  +\frac{1}{2}m^{2}\left(  \frac{1}{16\pi^{2}}\allowbreak \left(  M^{2}%
-m^{2}\right)  \left(  \gamma-1-\ln4\pi+\ln t\right)  \right)  .\nonumber
\end{align}
Or,%

\begin{align*}
\frac{E}{m^{4}}  &  =\frac{1}{2}\left(  1-\frac{12g}{4!}\frac{1}{16\pi^{2}%
}\allowbreak \left(  t-1\right)  \left(  \gamma-1-\ln4\pi+\ln t\right)
\right)  \frac{b^{2}}{\left(  16\pi^{2}\right)  }\\
&  -\frac{g}{4!}\left(  \frac{b}{4\pi}\right)  ^{4}+\left(  -\frac{1}%
{64\pi^{2}}\left(  t^{2}-1\right)  \left(  \gamma+\ln t-\ln4\pi-1\allowbreak
\allowbreak \allowbreak \right)  \right) \\
&  -\frac{3g\left(  \frac{1}{16\pi^{2}}\allowbreak \left(  t-1\right)  \left(
\gamma-1-\ln4\pi+\ln t\right)  \right)  ^{2}}{4!}\\
&  +\frac{1}{2}\left(  \frac{1}{16\pi^{2}}\allowbreak \left(  t-1\right)
\left(  \gamma-1-\ln4\pi+\ln t\right)  \right) ,
\end{align*}
where we used the dimensionless parameter $b$ such that $B=\frac{b}{4\pi
m^{2}}$. Also, in using dimensionless quantities of the form $e=\frac
{32\pi^{2}}{m^{4}}E$ and $g=(4\pi)^{2}G$, we have the dimensionless vacuum
energy of the form;%

\begin{align}
e  &  =-\left(  \frac{1}{2}G\left(  t-1\right)  \left(  \gamma+\ln t-\ln
4\pi-1\right)  -1\right)  \allowbreak b^{2}\nonumber \\
&  -\frac{G}{12}b^{4}+\left(  -\frac{1}{2}\left(  t^{2}-1\right)  \left(
\gamma+\ln t-\ln4\pi-1\allowbreak \allowbreak \allowbreak \right)  \right)
\label{edimls}\\
&  -\frac{G\left(  \allowbreak \left(  t-1\right)  \left(  \gamma-1-\ln4\pi+\ln
t\right)  \right)  ^{2}}{4}\nonumber \\
&  +\left(  \left(  t-1\right)  \left(  \gamma-1-\ln4\pi+\ln t\right)
\right)  .\nonumber
\end{align}
To understand well the features of this result, let us note that the the
effective potential is the generating functional of the one-particle
irreducible (1PI) amplitudes \cite{Peskin}. This fact results in the relation;%

\begin{equation}
\frac{\partial^{n}}{\partial b^{n}}E(b,t,G)=g_{n}\text{,} \label{stab}%
\end{equation}
where $g_{n}$ is related to the n-point Greens function. For instance, the
two-point function can be generated from the effective potential via the
second derivative of the effective potential with respect to the condensate,
\textit{i.e},
\begin{equation}
\text{\  \ }\frac{\partial^{2}E}{\partial B^{2}}=-iD^{-1}=M^{2}\text{,}
\label{ren}%
\end{equation}
where $D$ is the propagator. Since $B$ does not depend on the position (zero
momentum) we have $D=i/(p^{2}-M^{2})=-i/M^{2}$. Thus, $g_{2}=M^{2}=-\frac
{1}{2}gB^{2}+m^{2}-\frac{1}{2}g\Delta$. Since in our work $M^{2}$ is
constrained to be positive, this shows that the propagator has the correct
sign (no ghosts). This unexpected result has been explained by Jones
\textit{et.al }as the mean field approach in $\mathcal{PT}$-symmetric theories
knows about the metric.

To analyze our results, we note that the stability condition $\frac{\partial
E}{\partial B}=0$ enforces $\bar{H}_{1}$ to be zero. Accordingly, we get the results;%

\begin{align}
\frac{1}{3}b\left(  -Gb^{2}+3G\left(  t-1\right)  \left(  -\gamma+2\ln2+\ln
\pi+1-\ln t\right)  +6\right)  \allowbreak &  =0,\nonumber \\
\allowbreak G\left(  t-1\right)  \left(  -\gamma+2\ln2+\ln \pi+1-\ln t\right)
+2-Gb^{2}  &  =2t \label{bt}%
\end{align}
From these equations one can get the results;
\begin{equation}
b=\sqrt{-\frac{3t}{G}}, \label{cond}%
\end{equation}%
\begin{equation}
2G\left(  t-1\right)  \left(  \gamma-2\ln2-\ln \pi+\ln t-1\right)  -4=2t.
\label{t4d}%
\end{equation}
Note that these results shows that the vacuum condensate predicted from
Eq.(\ref{cond}) is imaginary and thus the Hamiltonian in Eq.(\ref{quasi}) is
non-Hermitian but $\mathcal{PT}$-symmetric.

\begin{figure}[ptbh]
\begin{center}
\includegraphics{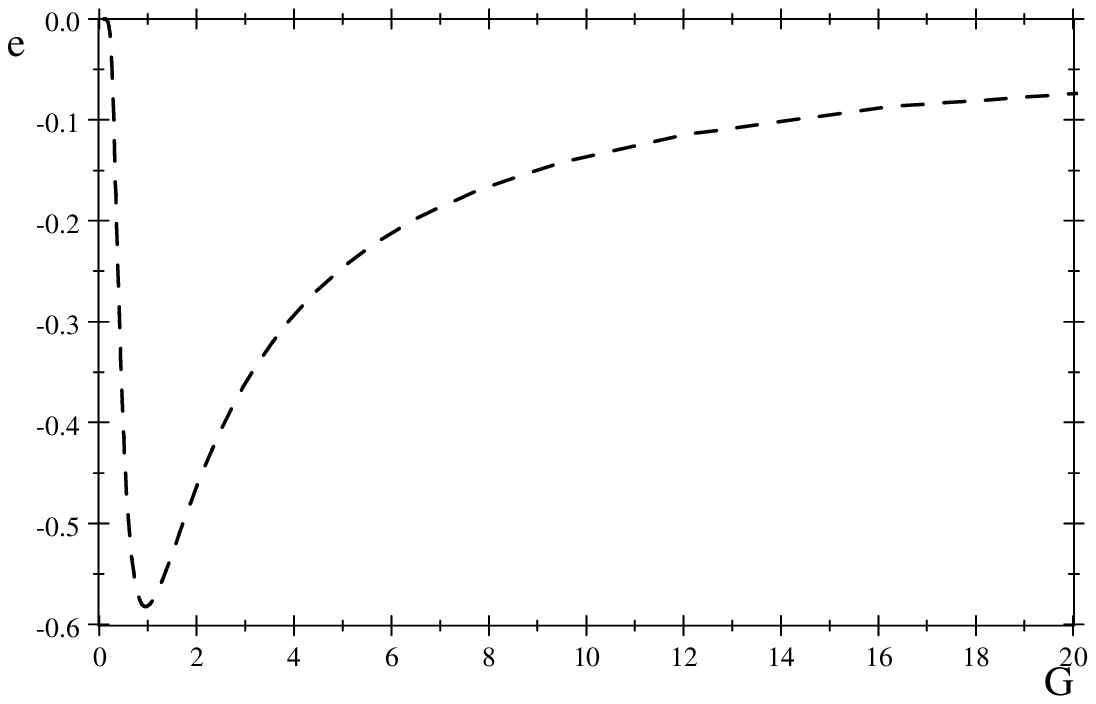}
\end{center}
\caption{The vacuum energy ( has a mass dimensions of 4) of the $(-\phi^{4})$
scalar field theory. For either $G$ small (high energy scales) or large (IR
energy scales) the vacuum energy is finite.}%
\label{en4d}%
\end{figure}

\begin{figure}[ptb]
\centering
\includegraphics{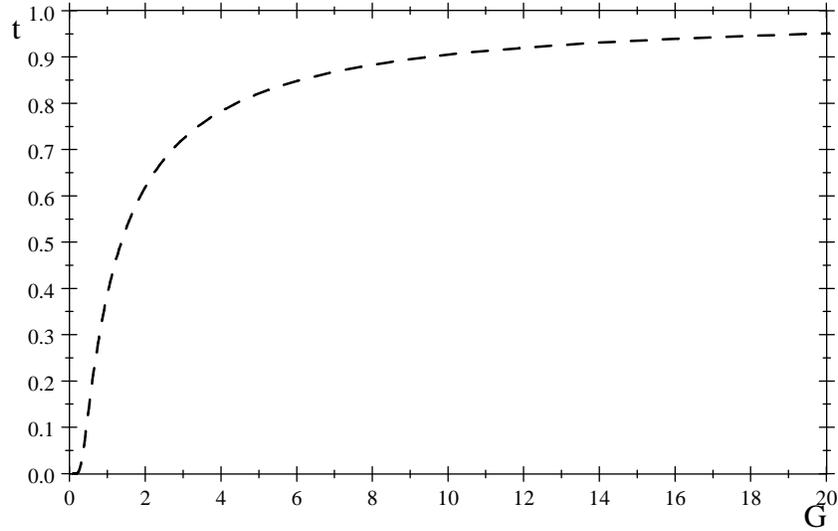} \caption{The mass squared $t$ ($t=\frac{M^{2}}{m^{2}})$ of the $(-\phi^{4})_{3+1}$ scalar field theory which has  a mass dimensions of 2. In this case also for either $G$ small (high energy scales) or large (IR energy scales) $t$ is
finite.}%
\label{t4dg}%
\end{figure}

\begin{figure}[ptb]
\centering
\includegraphics{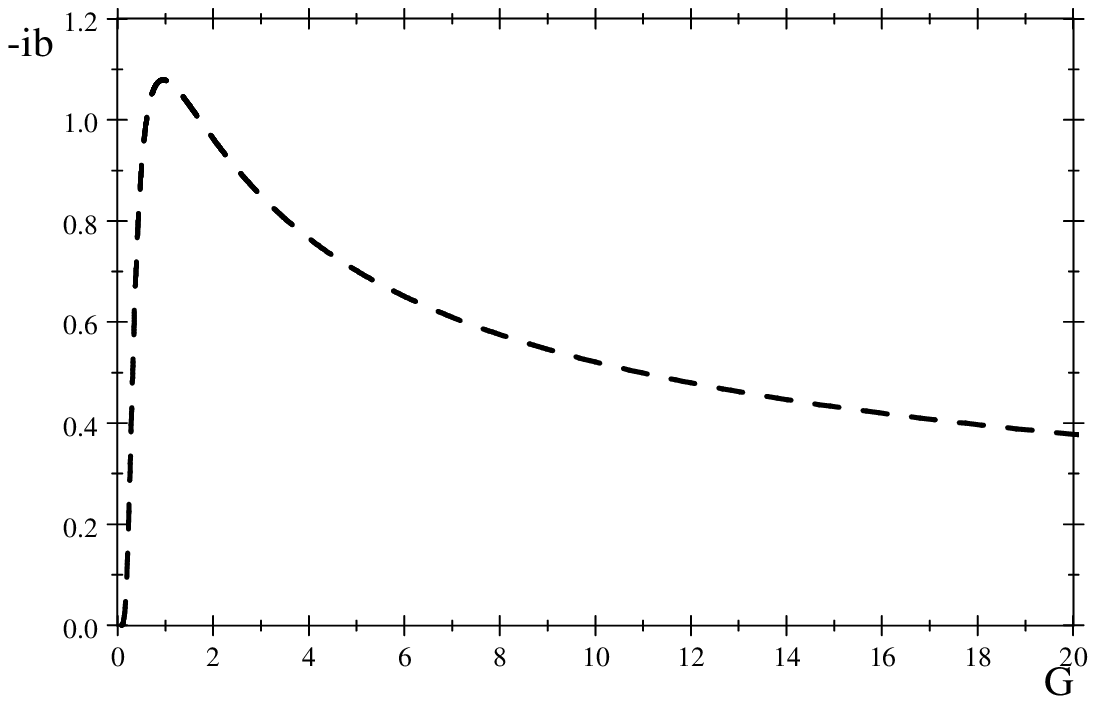} \caption{The absolute value of the vacuum condensate
( has a mass dimensions of 1) $\left|  b\right|  $ of the $(-\phi^{4})$ scalar
field theory. For either $G$ small (high energy scales) or large (IR energy
scales) the vacuum condensate is finite too.}%
\label{b4d}%
\end{figure}

One can solve Eq(\ref{t4d}) for $t$ as a function of $G$ and thus can obtain
the dependence of the vacuum energy $e$ and the vacuum condensate $b$ on the
coupling $G$. In Figs.\ref{en4d}, \ref{t4dg} and \ref{b4d} we show the
calculations of the vacuum energy, Higgs mass squared and the vacuum
condensate , respectively, as a function of the coupling $G$. Before w go on,
let us check the accuracy of our calculations. In Ref. \cite{bendg1}, Bender
\textit{et.al} obtained the behavior of the one point function $b$ as
$G\rightarrow0^{+}$ (numerically). They showed that $b$ goes to zero as
$\exp \left(  -\frac{\Lambda}{\epsilon}\right)  $, where $\Lambda$ is called
the controlling factor and $\epsilon$ there is related to the coupling. In
$3+1$ space-time dimensions they obtained the numerical value $\Lambda
=26.3209$. Now, Eq(\ref{t4d}) can be rewritten as
\[
G=\frac{t+2}{\left(  t-1\right)  \ln \frac{t}{c}},
\]
where $c$ is given by: $c=\exp \left( -\left( \gamma-2\ln2-\ln \pi-1\right)
\right) $. Accordingly, as the parameter $t\rightarrow0^{+}$ the coupling
$G\rightarrow0^{+}$. Now writing Eq.(\ref{t4d}) in the form;%
\[
G\left(  \gamma-2\ln2-\ln \pi+\ln t-1\right)  =\frac{t+2}{\left(  t-1\right)
},
\]
and since as $G\rightarrow0^{+}$ we have $t\rightarrow0$ then at this limit one
can approximate this equation by%

\begin{align*}
G\left(  \gamma-2\ln2-\ln \pi+\ln t-1\right)   &  \approx-\left(  t+2\right)
\left(  1+t\right)  ,\\
&  \approx-2-3t,
\end{align*}
which can be solved to give
\[
t=\frac{1}{3}G\omega \left(  \frac{3}{G}e^{-\frac{Gc+2}{G}}\right)  ,
\]
with $\omega \left(  x\right)  $ is the Lambert $\omega$ function defined by
$\omega \left(  x\right)  e^{\omega \left(  x\right)  }=x$. Note that, for small
arguments $\omega \left(  x\right)  \approx x$ and thus as $G\rightarrow0^{+}$,
the parameter $t$ can be approximated by;%

\[
t=\exp \left(  -c\right)  \exp \left(  \frac{-2}{G}\right)  ,
\]
and in using Eq.(\ref{cond}), we obtain the asymptotic behavior of the one
point function as $G\rightarrow0^{+}$ in the form
\begin{align*}
b  &  =\sqrt{-\frac{3\exp \left(  -c\right)  \exp \left(  \frac{-2}{G}\right)
}{G}}\\
&  =\pm i\sqrt{\frac{3}{G}}e^{-\frac{1}{2}c}e^{\frac{-1}{G}},
\end{align*}
This shows that we were able to obtain the exponential behavior for the condensate (analytically and for the first time) predicted numerically in Ref. \cite{bendg1}. Moreover, in accounting for the different
coupling used in our work from that in Ref.\cite{bendg1}, we find that the
controlling factor is given by $\Lambda=\frac{(4\pi)^{2}}{6}=\allowbreak26.\,
\allowbreak319$ compared to the numerical prediction of $\Lambda$ in Ref.
\cite{bendg1} as $26.3209$. This result assures the reliability of our
analytic calculations.

The result $t\rightarrow0^{+}$ as $G\rightarrow0^{+}$ obtained above seems to
be strange as one expects that $t\rightarrow1$ (no quantum corrections) at
this limit. To explain this result, we mention that, up to the first order
correction, $G$ here is the normalized coupling and if the theory is
asymptotically free it means that at the $UV$ scales $G\rightarrow0$.
Accordingly, at this limit all quantities of positive mass dimensions goes to
zero as well. Accordingly, the result $t\rightarrow0$ as $G\rightarrow0^{+}$
coincides with the spirit of the asymptotic freedom property concluded from
the real-line perturbative calculations. In other words, Eq.(\ref{t4d}) agrees
with the renormalization group flow of the coupling, small values of the
coupling correspond to high energy scales and vice versa. While quantities
that have positive dimension in terms of mass unit for the Hermitian $\phi
^{4}$ theory blow up at high energy scales (large coupling in this case), the
corresponding quantities in $\mathcal{PT}$-symmetric $(-\phi^{4})$ theory tend
to tiny values at high energy scales (small coupling in this case). This
realization pushes us to believe that the employment of the $\mathcal{PT}%
$-symmetric Higgs mechanism in the standard model may solve the famous
Hierarchy problem. These interesting features are appearing in our
calculations presented in Figs.\ref{en4d}, \ref{t4dg} and \ref{b4d}. For
instance, in Fig. \ref{en4d}, we plotted the vacuum energy which has a mass
dimension of $4$ and it is clear that the vacuum energy goes to zero as
$G\rightarrow0^{+}$. This is a very important result because in the
corresponding Hermitian theory, vacuum energy represents the worst case of the
Hierarchy problem which introduces the cosmological constant problem. In view
of our analysis, we strongly believe that the employment of the $\mathcal{PT}%
$-symmetric Higgs mechanism will solve the cosmological constant problem too.
Note that, finding it as a fully acceptable calculational algorithm for the
$\mathcal{PT}$-symmetric field theory \cite{aboeff,jonesgr,jonesgr1}, the mean
field approach is a systematic setup to handle the employment of the
$\mathcal{PT}$-symmetric Higgs mechanism in a complete manner.

In Fig. \ref{t4dg} \ a quantity of mass dimension $2$ is represented and also
assures the existence of the asymptotic freedom in the$\mathcal{PT}$-symmetric
$(-\phi^{4})$ field theory. 

In Fig. \ref{b4d}, the vacuum condensate which has a mass dimension of $1$ has
been plotted and assures the asymptotic freedom property too.

Let us now speculate about the actual case in the standard model where the
Higgs mass dominant contribution has the form \cite{DJOUADI};
\begin{equation}
M_{H}^{2}=M_{0}^{2}-\frac{\Lambda^{2}}{8\pi^{2}v^{2}}[M_{H}^{2}+2M_{w}%
^{2}+M_{Z}^{2}-4M_{t}^{2}],
\end{equation}
with the different parameters are defined in Ref.\cite{DJOUADI} ($\Lambda$
here represents a momentum cut off) and since all the species in the standard
model attain their masses from the vacuum condensate, thus the masses have
even sharper than an exponential decrease in $\Lambda$ \footnote{The coupling
is decreasing as a function of the energy scale as shown in Ref.\cite{abohir}%
.}. Hence, one expect that all the particles in the standard model (including
Higgs) have finite masses at high energy scales \footnote{The different group
structure between theory investigated in this work and the one suitable to
break the $SU(2)\times U(1)$ symmetry in the standard model will affect the
numerical values but not the main features of the theory and thus conclusions
are supposed to be the same in both cases.}.

To conclude, we have calculated the vacuum energy for the non-Hermitian and
$\mathcal{PT}$-symmetric $(-\phi^{4})$ scalar field theory in $3+1$
dimensions. We find that the vacuum energy is small for the whole range of
energy scales which is a very interesting sign that this theory is a very good
candidate to play the role of the Higgs mechanism in the standard model of
particle interactions. We assert that the conventional Hermitian $(\phi^{4})$
has a vacuum energy which blows up at high energy scales as a manifestation of
the famous Hierarchy puzzle. In fact, this adds a large value to the
cosmological constant and thus enhancing the unacceptable discrepancy between
theoretical and experimental predications of the cosmological constant.
Accordingly, the discussions here supports the fruitfulness of replacing the
Hermitian Higgs mechanism by the $\mathcal{PT}$-symmetric $(-\phi^{4})$ one.

Since in the literature the asymptotic freedom is always assumed for the
theory under consideration by the aid of real-line calculations, we generated
the diagrams in Figs.\ref{en4d}, \ref{t4dg} and \ref{b4d} using our effective
field calculations. The effective Hamiltonian is a real-line theory and thus
conclusions from our calculations are then reliable. In fact, the quantities
generated in the graphs have positive mass dimensions and according to
asymptotic freedom they have to vanish at very high energy scales (small
coupling), which is very clear from the figures. While the naive perturbation
analysis in Ref.\cite{abohir} shows the same result, the Higgs mass blows up
at small energy scales (large coupling). However, at large couplings
perturbations are meaningless and conclusions have to be drawn from
non-perturbative treatments of the theory. In fact, the non-perturbative
effective field approach we used cured this problem as we can realize from
Fig.\ref{t4dg} that the mass parameter is finite for the whole energy range
(or the whole range of the coupling $G$).

The effective field approach used in this work also knows about the metric as
one can realize that the propagator has its correct sign. This prediction has
recently proved by Jones \textit{et.al} in Refs.\cite{jonesgr,jonesgr1}.
Accordingly, there exist no need to take care about the so far unattained
metric operator. Thus, the algorithm we followed here is sufficient to tackle
the theory and can be employed easily to the realistic case of the standard
model. Moreover, the accuracy of the algorithm has been tested in a quantitative manner in
Ref.\cite{aboeff} and it was found that the effective field approach is
reliable for the study of $\mathcal{PT}$-symmetric theories.

To check the validity of our calculations, we obtained the asymptotic behavior
of the vacuum condensate as a function of the coupling when $G\rightarrow
0^{+}$. At this limit, we found that the condensate behaves like $\exp
(\frac{-1}{G})$ a result which was conjectured by bender \textit{et.al} in
Ref.\cite{bendg1}. Moreover, we obtained the controlling factor $\Lambda
=\frac{(4\pi)^{2}}{6}$ which is very close to the numerical prediction
obtained by bender \textit{et.al} in Ref.\cite{bendg1}.

\begin{acknowledgments}
We would like to thank Dr. S.A. Elwakil for his support.
\end{acknowledgments}

\newpage

\newpage 

\begin{thebibliography}{99}                                                                                               %


\bibitem {DJOUADI}Abdelhak DJOUADI, Phys.Rept.457:1-216 (2008).

\bibitem {abohir}Abouzeid shalaby, Will the PT Symmetric and Non-Hermitian
$\phi^{4}$ Theory Solve the Hierarchy and Triviality Problems in the Standard
Model, arXiv:0712.2521.

\bibitem {bendh1}Bender and Milton, Phys. Rev. D 55, R3255 (1997).

\bibitem {abophi41p1}Abouzeid shalaby, Eur.Phys.J.C50:999-1006 (2007).

\bibitem {abophi61p1}Abouzeid shalaby, Phys.Rev.D76:041702 (2007 ).

\bibitem {zadah}A. Mostafazadeh, J. Phys. A: Math. Gen. 38, 6557 (2005) and
Erratum 38, 8185 (2005).

\bibitem {bendr}Carl Bender and Stefan Boettcher, Phys.Rev.Lett.80:5243-5246 (1998).

\bibitem {bend2005}Carl M. Bender, Jun-Hua Chen and Kimball A. Milton,
J.Phys.A39, 1657 (2006).

\bibitem {bendg1}Carl M. Bender, Peter N. Meisinger, and Haitang Yang, Phys.
Rev. D63, 045001 (2001).

\bibitem {Symanzik}K. Symanzik, Commun. Math. Phys. 45, 79 (1975).

\bibitem {bendf}C. M. Bender, K. A. Milton, and V. M. Savage, Phys. Rev. D 62,
85001 (2000).

\bibitem {Frieder}Frieder Kleefeld, J. Phys. A: Math. Gen. 39 L9--L15 (2006).
\bibitem {Peskin}Michael E. Peskin and Daniel V.Schroeder, An Introduction To Quantum Field Theory (Addison-Wesley Advanced Book Program) ( 1995).

\bibitem {kaku}M. Kaku, Quantum Field Theory, Oxford University Press, Inc.(1993).

\bibitem {susy}{J. Wess and B. Zumino, Nucl. Phys. B 70, 39 (1974) .}


\bibitem {tecni}{L. Susskind, Phys. Rev. D 20, 2619 (1979) .}

\bibitem {tecni1}{M. E. Peskin and T. Takeuchi, Phys. Rev. Lett. 65, 964 (1990); M. Golden and L. Randall, Nucl. Phys. B 361, 3 (1991); B. Holdom and J.
Terning, Phys. Lett. B 247, 88 (1990).}

\bibitem {extra}Yutaka Sakamura, Phys.Rev.D76, 065002 (2007).

\bibitem {aboeff}Abouzeid Shalaby, Phys. Rev. D 79, 065017 (2009).

\bibitem {quart}Jing-Ling Chen, L.C. Kwek and C.H.Oh, Phys. Rev. A 67, 012101 (2003).

\bibitem {realline}Carl M. Bender, Dorje C. Brody and Hugh F. Jones,
Phys.Rev.D73, 025002 (2006).

\bibitem {jonesgr}H. F. Jones, hep-th/1002.2877.

\bibitem {jonesgr1}H.F. Jones and R.J. Rivers, Phys. Let. A 373, 3304-3308 (2009).

\bibitem {Coleman}S.Coleman, Phys.Rev. D11, 2088 (1975).

\bibitem {efbook}M.~Dineykhan, G.~V.~Efimov, G.~Ganbold and S.~N.~Nedelko,
Lect.\ Notes Phys.\  \textbf{M26}, 1 (1995).

\bibitem {wen-Fa}Wen-Fa Lu and Chul Koo Kim, J. Phys. A: Math. Gen. 35 393-400 (2000).

\bibitem {changcac}Chang S. J., Phys. Rev. D 12, 1071 (1975).

\bibitem {mag}Steven. F. Magruder, Phys.Rev. D 14,1602(1976).

\bibitem {normal2}Allan M. Din, Phys.Rev.D4,995 (1971).

\bibitem {collinbook}Book by John C. Collins "Renormalization", Cambridge University Press (1984).

\end{thebibliography}
\end{document}